\documentclass[preprint2]{aastex}
\usepackage{graphics}
\usepackage{epsfig}

\def\chandra{{\sl Chandra}}

\def\g359{{G359.89--0.08}}
\def\G359{{G359.89--0.08}}

\def\xs{\hbox{G359}}

\def\sgra{\hbox{Sgr~A$^*$}}
\newcommand{\as}{$^{\prime\prime}$}

\newcommand{\etal}{et al.~}

\begin{document}

\slugcomment{draft version of Oct 24 2005}
\shorttitle{G359.95-0.04: Pulsar Candidate Near Sgr~A$^*$}
\shortauthors{Wang et al.}

\title{G359.95-0.04: An Energetic Pulsar Candidate Near Sgr~A$^*$}

\author{Q.~D. Wang\altaffilmark{1,2}, F.~J. Lu\altaffilmark{1,3}, and E.~V. Gotthelf~\altaffilmark{4}}
\altaffiltext{1}{Astronomy Department, University of Massachusetts, Amherst, MA 01003; wqd@astro.umass.edu; lufj@astro.umass.edu}
\altaffiltext{2}{Institute for Advanced Study, Einstein Drive, Princeton, NJ 08540}
\altaffiltext{3}{Laboratory of Particle Astrophysics, Institute of High Energy Physics, CAS, Beijing 100039, P.R. China}
\altaffiltext{4}{Columbia Astrophysics Laboratory, Columbia University, 550 West 120th Street, New York, NY 10027; eric@astro.columbia.edu}

\begin{abstract}
We report the discovery of a prominent nonthermal X-ray feature
located near the Galactic center that we identify as an energetic
pulsar wind nebula. This feature, G359.95-0.04, 
lies 1 lyr north of \sgra\
(in projection), is comet-like in shape, and has a power law spectrum
that steepens with increasing distance from the putative pulsar. The
distinct spectral and spatial X-ray characteristics of the feature are
similar to those belonging the rare class of ram-pressure confined
pulsar wind nebulae. The luminosity of the nebula
at the distance of \sgra, consistent with the inferred X-ray
absorptions, is $L_x \sim 1 \times 10^{34} {\rm~ergs~s^{-1}}$ in the
2--10 keV energy band.  The cometary tail extends back to a region
centered at the massive stellar complex IRS 13 and surrounded by enhanced 
diffuse X-ray emission, which may represent an
associated supernova remnant. Furthermore, the inverse Compton 
scattering of the strong ambient radiation by the nebula 
consistently explains the observed TeV emission from 
the Galactic center. We also briefly discuss plausible connections
of G359.95-0.04 to other high-energy sources in the region, such as the 
young stellar complex IRS 13 and SNR Sgr A East. 

\end{abstract}


\section{Introduction}

The Galactic center (GC) provides a unique laboratory for a detailed
study of the interplay between massive star formation and galactic
nuclear environment.  Particularly interesting is the presence of a
young massive stellar cluster around Sgr A$^*$ --- the super-massive
black hole (SMBH) of the Galaxy (e.g., \citet{gen03}). This
cluster, with an integrated mass of $\sim 10^4 M_\odot$, has an age of
$\sim 6 \times 10^6$ yrs and a flat initial mass function; therefore a
considerable number of supernovae (SNe) should have occurred in this
region. The presence of their stellar end-products (e.g., pulsars) and
supernova remnants (SNRs) could be responsible for various high-energy
phenomena observed in the GC region and could also have strong
impacts on the environment, and hence on accretion onto the SMBH (e.g.,
\citet{aha04, bel05}; for a review, see \citet{meli01}).

However, no pulsar has yet been found within $\sim 1^\circ$ of the
Galactic center.  Traditional radio searches at $\lesssim 1$ GHz are
insensitive to pulsars in this region because interstellar scattering
causes severe pulse broadening \citep{cor97}.  At high
frequencies ($\nu \gtrsim 10$ GHz), where pulse broadening ($\propto
\nu^{-4}$) is not as important, blind pulsar searches are difficult
because radio telescope beams are small
and because pulsar spectra are typically steep, with correspondingly less
flux available.
More productive are searches in the hard ($
\gtrsim 2$ keV) X-ray band, where young
pulsars, their wind nebulae (PWNe), and/or SNRs can be
identified by X-ray imaging. Indeed, strong X-ray emission from SNR
Sgr~A~East, near Sgr A$^*$, has long been known. In particular, \citep{par05} 
have suggested that a point-like hard X-ray source
CXOGC~J174545.5--285829, or ``cannonball'', is a candidate of a young
pulsar, which may have been ejected from the SNR.

Recently, from the larger GC region \citep{wan02a}, 
\citet{wan02} and \citet{lu03} 
have detected three X-ray threads associated with
nonthermal radio filaments or radio ``wisps''. These highly polarized
radio features, observed only in the GC, are due to synchrotron
radiation from relativistic particles (electrons and/or positrons).  
The detection of the X-ray counterparts to these filaments provides strong
constraints on the particle acceleration mechanism.  The X-ray
emission, if also due to the synchrotron process, must arise from
particles accelerated nearly {\sl in situ}. A possible origin of these
particles is PWNe, perhaps shaped by the strong magnetic field and/or
ram-pressure of the GC \citep{wan02}. Alternatively, some of the
X-ray threads may represent shocks of young SNRs \citep{ho85, sak03}.

Here we report on our study of a prominent nonthermal X-ray feature 
which probably represents the most convincing case for the
presence of a young pulsar in
the GC. This pulsar is separated from \sgra\ by
only $\sim$ 8\farcs7 at its distance of $8$ kpc, corresponding to a
projected separation of 0.32~pc. 
In the following, all error bars of our X-ray measurements are
presented at the 90\% confidence level.

\section{Data Calibration}

This work takes advantage of a large number of {\sl Chandra} 
observations aimed at \sgra. We utilize the archival data 
available by September 2005, including
the twelve observations taken before May 3, 2002, as 
listed by \citet{par05}, and three more recent ones 
(Obs. \# 3549, 4683, and 4684) taken on June 19, 2003 and July 5-6, 2004. 
The total resultant exposure of the included observations is 710 ks. 
We first process individual observations, following standard CIAO 
(version 3.2.2) event reprocessing procedures, including the 
correction for charge transfer inefficiency, 
bad-pixel-removing, and light-curve cleaning.
We corrected the absolute astrometry 
of each observation by matching the centroid of the X-ray source 
CXOGC~J174540.03--290028.2 
with its counterpart (Sgr A$^*$) position [R.A.,
  Dec. (J2000) $= 17^h45^m40^s.041, -29^\circ,0^{\prime} 28\farcs12$;
\citet{rei99}]
obtained in the high resolution (75~mas) near-IR (SINFONI) image \citep{eis05}. 
While CXOGC~J174540.03--290028.2 is clearly resolved, we use the radial
surface intensity of the source CXOGC~J174540.0--290031, a low-mass
X-ray binary (LMXB) with an eight hour orbital period 
\citep{mun05},  to characterize the point spread
function, which is compared to other nearby sources to look for
possible deviations from point-like emission. The astrometry uncertainty and 
spatial resolution of the
final, merged data set are $\sim 0\farcs15$ and 0\farcs9 (FWHM),
respectively, for the region of interested here (Fig.~\ref{ims}). Part of the PSF
broadening is due to the dust scattering along the sight line to the GC
\citep{tan04}.

We construct event images with a bin size of 0\farcs0984 in the
1--2.5, 2.5--4, 4--6, 6--9 keV bands. The corresponding exposure-corrected
(flat-fielded) images are constructed with using a weighted spectrum
assuming an absorbed power law model with a photon index 1.7 and
foreground cool gas column of N$_H = 10^{23}$~cm$^{-2}$. The
flat-field correction is small ($\lesssim 10\%$) over our region of
interest.

\section{Analysis and Results}

\begin{figure*}[thb!]
 \centerline{
\psfig{figure=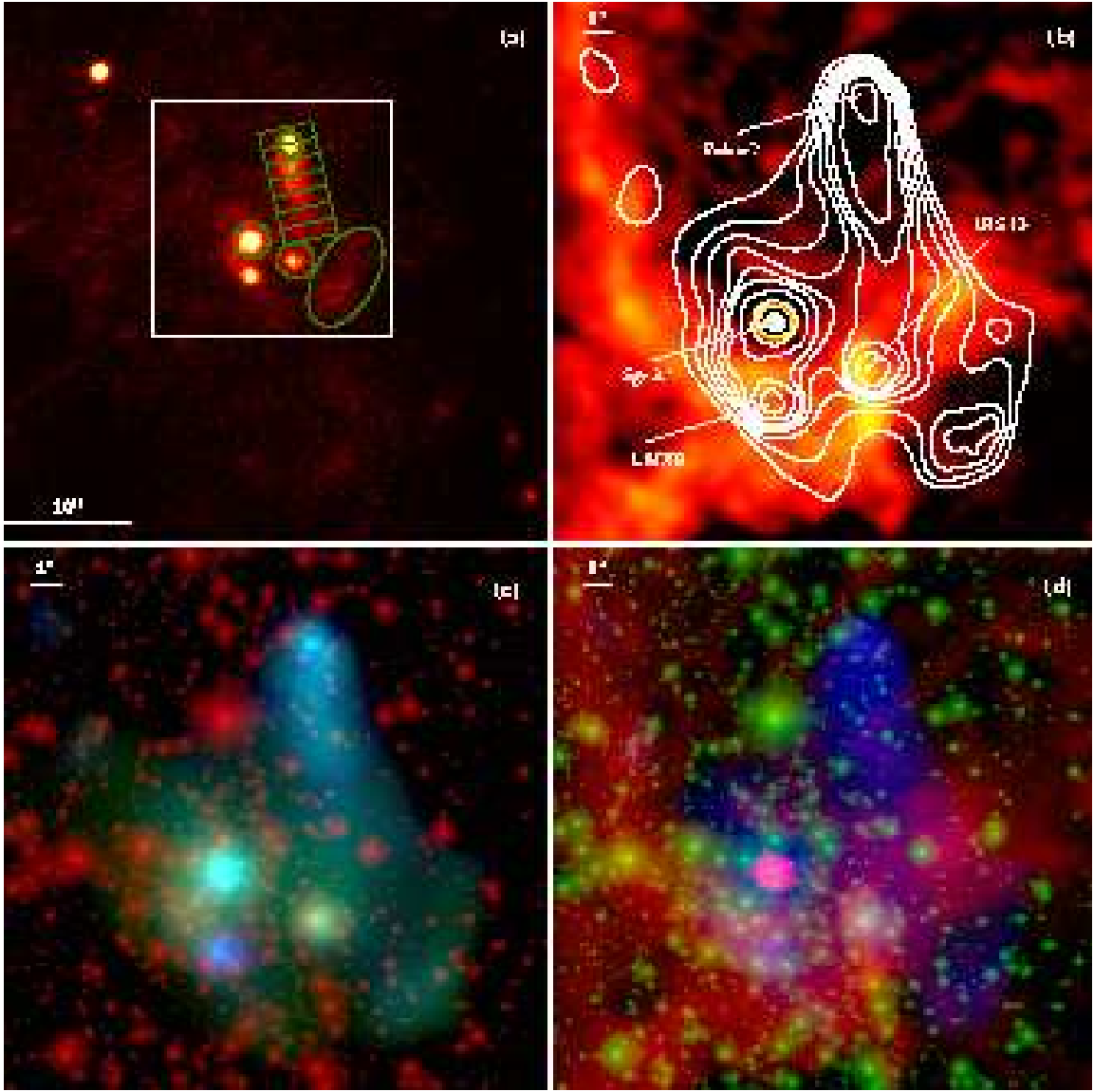,height=6.in,angle=0, clip=}
}
\caption{The immediate vicinity of \sgra\ in radio, near-IR, 
and X-ray (North is up and east is to the left):
(a) ACIS-I count image of the field ( $40$\as\
$\times$ 40\as) around \sgra\ in the 1--9~keV
band. The image is smoothed with a Gaussian with a FWHM of 0\farcs2.
Various regions for spectral extractions are marked (see text for details).
The largest square box outlines the field covered by the images 
shown in (b)-(d).
(b) 3.6 cm continuum image \citep{rob93} and {\sl Chandra} 
ACIS-I 1--9~keV  intensity contours at 0.48, 0.57, 0.69, 0.84,  1.0,
      1.2, 1.8, 3.3, and 6.3 ${\rm~counts~s^{-1}~arcmin^{-2}}$.
The X-ray intensity map is smoothed adaptively
(using the CIAO {\sl csmooth} routine) to achieve $S/N \sim 3$. 
(c) Tri-color presentation: the SINFONI near-IR image 
(red; \citet{eis05}), ACIS-I 2.5--4 keV image (green),
and ACIS-I 4--9 keV image (blue). The X-ray intensity images are smoothed
in the same way as in (b). (d) Tri-color presentation: the
radio continuum image (red; as in (b)), the near-IR image (green; as in (c)),
and the 2.5-9 keV image (blue; as in (c)).
\label{ims}}
\end{figure*}

Fig.~\ref{ims} shows the field around \sgra in radio, near-IR, and X-ray. 
There is little general correlation among features in these wavelength bands.
The obvious exceptions are Sgr A$^*$, showing strong emission in 
both radio and X-ray, and IRS 13 in all three bands.
Fig.~\ref{ims}a shows the large-scale environment of the central concentration 
of X-ray sources and diffuse features in the close vicinity of Sgr A$^*$.
There is an apparent (though relatively low surface brightness) patchy 
shell-like structure, on scales of $\sim 30^{\prime\prime}$,
especially to the northwest of Sgr A$^*$. Here we focus on
the central concentration, particularly the outstanding comet-shaped 
X-ray feature, referred to herein as G359.95-0.04, or \xs\ for short. This
feature is substantially more
elongated and brighter than the other X-ray-emitting PWN candidate,
``cannonball'', 2\farcm3 northeast of Sgr A$^*$ \citep{par05}. 
The X-ray emission from this latter source shows only a tiny
tail and has a luminosity of $3.1 \times 10^{33} {\rm~ergs~s^{-1}}$ in
the 2--10 keV band.  The front end of \xs\ 
was detected previously as CXOGC~J174539.7--290020
\citep{bag03} and is located at coordinates $17^h45^m39^s.80,
-29^\circ 00^{\prime} 19\farcs9$ in our constructed image (Fig.~\ref{ims}). 
This position is 8\farcs7
from \sgra, corresponding to a projected separation of 0.32~pc. The
tail of \xs\ extends to a region even closer to \sgra\ ($\sim
4^{\prime\prime}$ away).  Part of the tail was detected as
CXOGC~J174539.7--290022 \citep{bag03}.  The combined extent of
this source was mentioned by \citet{bag03}, although no
further analysis or discussion of its nature was given. Figs.~1-2
clearly show that \xs\ represents a coherent structure, $\sim 2$\as\
wide and $\sim 8$\as\ long.  The front end appears to be point-like, with a full
intrinsic extent $\lesssim 0\farcs3$. A comparison of the count
rates in individual observations shows no evidence for significant
variability, although we cannot rule out flux changes at the
$\lesssim 40\%$ level.  The mean count rate of this point-like source is $3.7
\times 10^{-3} {\rm~counts~s^{-1}}$, or $\sim 28\%$ of the rate from
the entire \xs\ feature.
  
\begin{figure*}[thb!]
 \centerline{
\psfig{figure=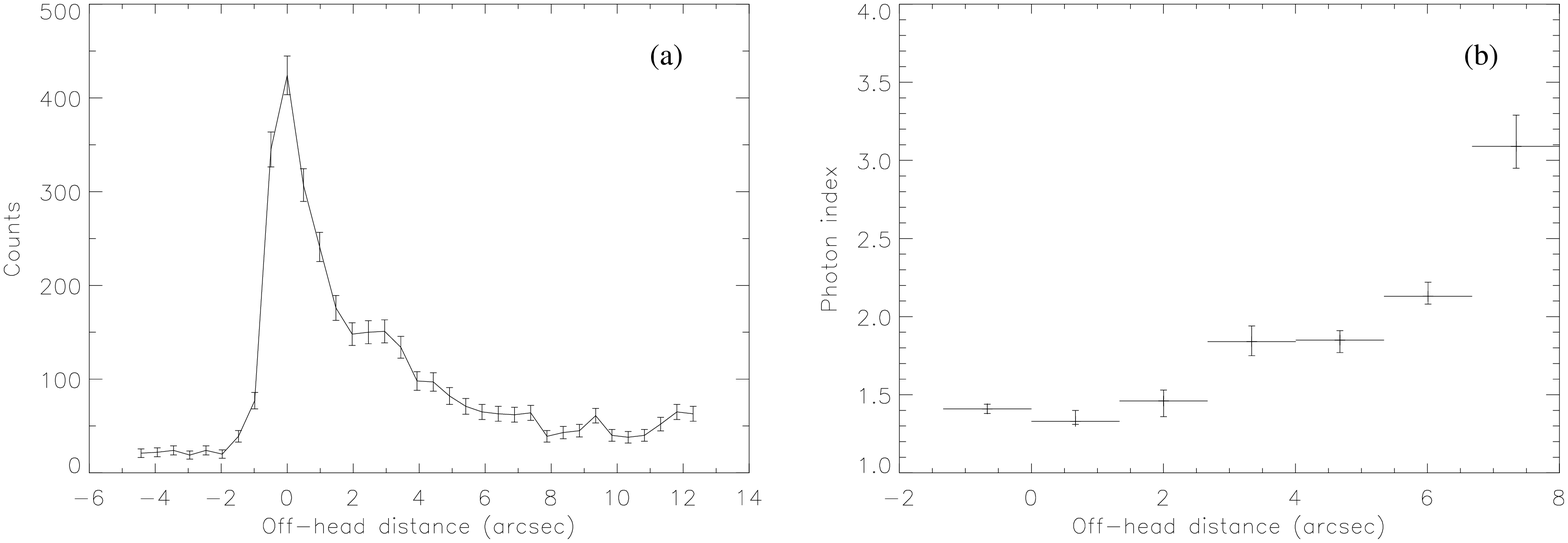,height=2.5in,angle=0, clip=}
}
\caption{ACIS-I count (a) and power law photon index (b) distributions
along the major axis of \xs; the positive distance is toward the
southwest of the point-like source centroid.
\label{prof}}
\end{figure*}

\begin{figure*}[thb!]
 \centerline{
\psfig{figure=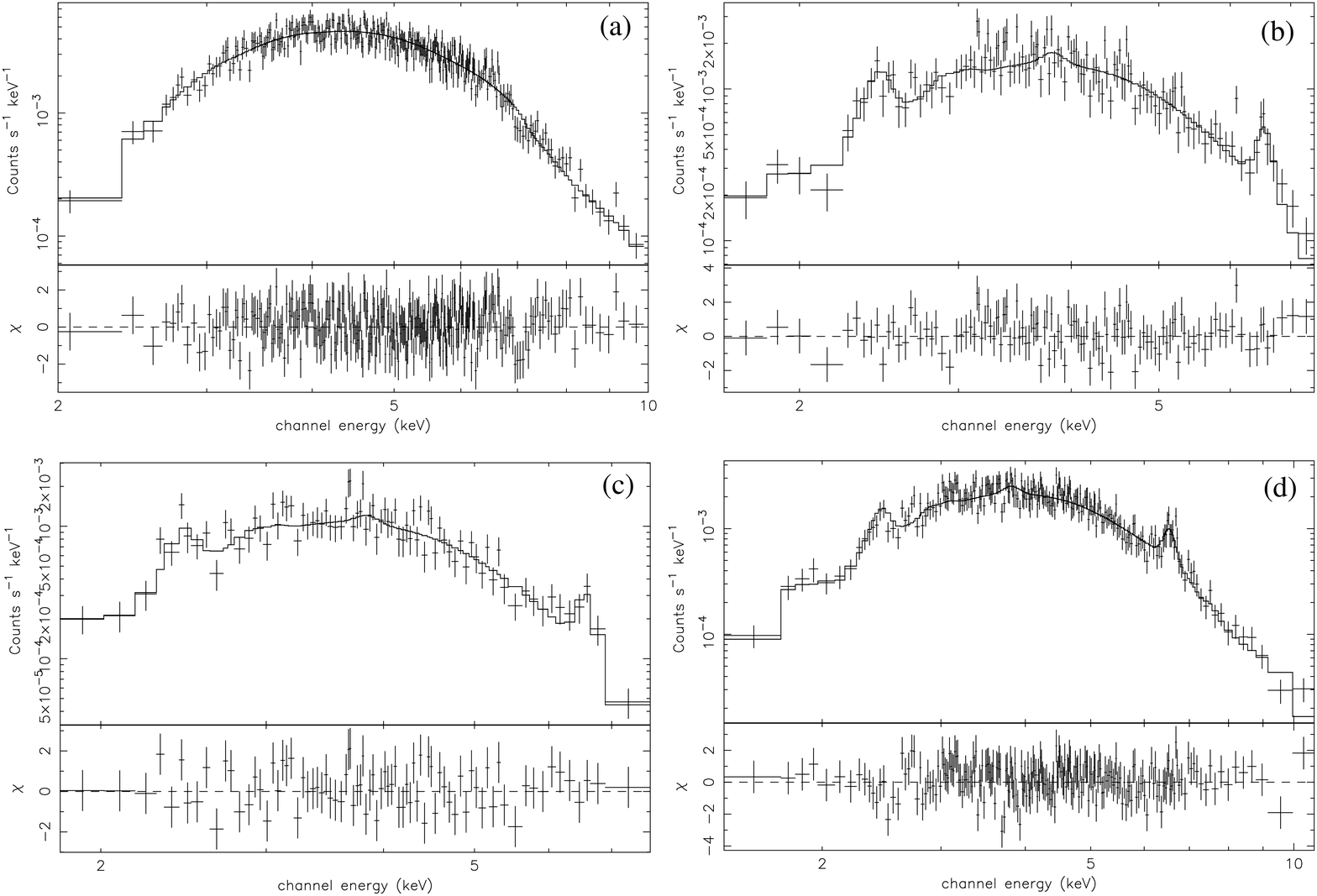,height=4.4in,angle=0, clip=}
}
\caption{{\sl Chandra} ACIS-I spectra of \xs\ (a),
SWXE (b), IRA 13 (c), and Sgr A$^*$ (d). The relative deviations 
from the best-fit models (Table~\ref{tab:spec}) are shown in the 
respective bottom panels.
\label{spec}}
\end{figure*}

Fig.~\ref{spec} presents ACIS-I spectral data and model fits of \xs\ and 
three relevant X-ray features in the region. These features 
are characterized mainly to facilitate X-ray spectral comparison
and discussion on their potential links to \xs\ (\S~6). 
The results of the fits are summarized in Table~\ref{tab:spec}.
\xs\ shows a hard and featureless X-ray spectrum (Fig.~\ref{spec}a).
We extract the ACIS-I spectra of the point-like source and the entire 
feature, using a 1\as\ radius aperture and a $9\farcs3 \times
4\farcs4$ rectangular box, as outlined in Fig.~\ref{ims}a. We further estimate
the local background from a field north of \xs; alternative background
fields are used to test the effect of the background choice on our
spectral results.  The overall energy spectrum of the entire \xs\  can be well
characterized by a power law, $f_x = 4.8 \times 10^{-4}  \epsilon^{-1.94}
{\rm~ph/(keV~s~cm^2)}$, which $\epsilon$ is the photon energy
in units of keV. Residuals to this spectral fit (the 
lower panel of Fig.~\ref{spec}a), however, shows an apparent dip, which can be 
characterized by a negative
Gaussian with a centroid of $7.04(6.96, 7.09)$~keV and an equivalent
width of $88(49, 136)$~eV. This dip, if real, is difficult to
explain. The energy centroid roughly corresponds to the H-like
Fe~K$\alpha$ line. But there is no evidence at $\sim 6.7$ keV for a
He-like Fe~K$\alpha$ absorption line, which should be stronger.

To examine the spectral change along the major axis of \xs, we divide
the rectangular box into seven sections (Fig.~\ref{ims}) and conduct a joint
power law fit to their spectra with a shared absorption column density
$N_H$. This gives a
satisfactory fit ($\chi^2/d.o.f. = 267/322$) for $N_H = 1.25 (1.09, 1.39) \times 10^{23} {\rm~cm}^{-2}$, with the photon index
ranging from $\Gamma \sim 1.4$ at the point-like source 
position to $\Gamma \sim
3.0$ towards the other end of comet-like structure (Fig.~\ref{prof}b). Note that the
fitted N$_{\rm H}$ does depend on the assumed spectral shape; the
systematic uncertainty in N$_{\rm H}$ is probably up to $\sim
20\%$. Fig.~\ref{prof}b clearly shows a systematic trend, with spectral
steepening that clearly increases with distance away from the point-like
source centroid,
although the last data point is sensitive to the background choice.

\begin{deluxetable}{lcccccccc}
\tablewidth{0pt}
\tablecaption{Spectral Fit Results\label{tab:spec}}
\tablehead{
Source & Model&$N_H$  & $\Gamma$  & $T$        &$n_e t$ &$\chi^2/d.o.f.$&$F_x$&$L_x$
}
\startdata
\xs\     &PW     & 1.43(1.34, 1.53)& 1.94(1.80, 2.11)& --- &---& 274/269&5.9 &10\\
SWXE     &GNEI   &1.03(0.94, 1.12)&---         &2.5(2.1, 3.0) & 2.9(2.0, 4.0) &120/114&1.2 & 2.7\\
IRS 13   &GNEI   &1.06(0.98, 1.20) & ---        &2.0(1.7, 2.3)  & 3.7(1.9, 4.4) &83/77 &0.73 &2.0\\
Sgr A$^*$&GNEI+PW&1.10(1.01, 1.20)&-1.4($\le$0.4)  &2.6(1.6, 3.4) & 3.0(2.3, 3.9) &211/209& 2.5&4.8\\
\enddata
\tablecomments{The included spectral models are power law (PW) and 
non-equlibrium ionization plasma (GNEI).  The key model parameters given are
the absorption column density $N_H$ in units of $10^{23}$~cm$^{-2}$,
the power law photon index $\Gamma$, 
the plasma temperature $T$ in keV, and the ionization parameter 
$n_e t$ in $10^{10}$~cm$^{-3}~s$. Also included are the observed 
flux $F_x$(2-10~keV) in $10^{-13} {\rm~ergs~s^{-1}~cm^{-2}}$ and the
absorption-corrected luminosity $L_x$(2-10~keV) in 
$10^{33} {\rm~ergs~s^{-1}}$ (a distance of $8$ kpc is assumed). Uncertainty intervals of the parameters are
all statistical at the 90\% confidence.}
\end{deluxetable}

G359 seems to emerge from a region near IRS 13, which is surrounded by
enhanced diffuse X-ray emission (Fig.~\ref{ims}). Part of this emission
from the southwest of CXOGC~J174540.0--290031, however, is due to
scattered flux from CXOGC~J174540.0--290031 \citep{mun05}. 
We extract a spectrum from the southwestern X-ray enhancement (SWXE; the 
region enclosed by the ellipse in Fig.\ref{ims}a), which 
is least confused with discrete sources. 
A local background is obtained from a
field further southwest. The background-subtracted diffuse X-ray
spectrum (Fig.~\ref{spec}b) shows two distinct line features at 
$\sim 2.44$~keV
and 6.55~keV, which are at the intermediate energies between the
neutral and He-like K$\alpha$ transitions of S and Fe, respectively. A
natural explanation of these line signatures is then 
that the diffuse X-ray emission arises from an optically-thin
thermal plasma in a non-equilibrium ionization (NEI) state. Indeed, 
the spectrum is reasonably well characterized by the XSPEC model {\sl gnei} with 
the abundance set to Solar (Table~\ref{tab:spec};
Fig.~\ref{spec}b). The ionization timescale averaged plasma temperature is not
constrained. We find that similar models (e.g., {\sl nei}, {\sl
npshock}) give essentially the same results.

The young stellar complex IRS 13 nearly
coincides with the centroid of an apparently resolved X-ray source
(CXOGC~J174539.7--290029; Fig.~\ref{ims}; \citet{bag03} 
and references therein). The slight offset ($\sim 0\farcs3$ to the west) 
of this X-ray source from IRS 13 is only marginally significant. 
Surprisingly, the ACIS-I spectrum of the source (Fig.~\ref{spec}c)
shows the same line signature for an NEI plasma as in the spectrum
of the SWXE (Fig.~\ref{spec}b). Similar signature is also indicated in a
spectrum of Sgr A$^*$ from a 51 ks ACIS-I 
observation \citep{bag03}. For comparison, we 
extract a spectrum for Sgr A$^*$ (Fig.~\ref{spec}d) 
from the present multiple ACIS-I
observations (Fig.~\ref{ims}a). This spectrum is remarkably similar
to that of IRS 13, clearly exhibiting the NEI line signature 
(Fig.~\ref{spec}c and d). Both spectra can be well fitted with the {\sl gnei} model.
The Sgr A$^*$ spectrum does show a significant excess above the model at high
energies ($\sim 8$ keV); a combination of the model with a power law 
further improves the fit (Fig.~\ref{spec}d; Table~\ref{tab:spec}). Although
such a spectral decomposition or even the choice of the {\sl gnei}
model may be far from being physical, the similar NEI nature of these 
X-ray-emitting objects is apparent. The absorption column densities 
toward these sources or features
are comparable, except perhaps for \xs\ (Table~\ref{tab:spec}). As shown
above, the higher $N_H$ value obtained in the spectral fit to \xs\ is 
at least partly due to the over-simplified power law description 
of this complex 
feature. Indeed, the 90\% $N_H$ lower limit from the above joint power law fit
to the divided spectra of the feature is consistent with the values
for the other sources. If the excess absorption is real, however, then \xs\  
may be embedded in a dense cloud and/or may
be significantly beyond Sgr A$^*$ in location. 

\section{G359.95-0.04 as a Ram-pressure-confined PWN}

Any interpretation of \xs\ needs to account for its distinct X-ray
characteristics: the cometary shape, the hard and apparently
nonthermal spectrum, the systematic spectral steepening with distance
from the front end, and the relatively high X-ray flux in contrast to
little emission in both radio and infrared. We find that
a ram-pressure confined PWN (e.g., \citet{wan93, van04}) provides 
a natural explanation of these X-ray characteristics. 
The term ``ram-pressure confinement'' is loosely defined here to mean
that a large pressure gradient is responsible for
producing the compact head and the one-side outflow morphology of  \xs. 
The region around Sgr A$^*$ may be
largely occupied by low-density hot gas (e.g., reverse shock heated SN ejecta), 
in which a weak 
bow shock or a pressure gradient confinement is likely formed upstream
\citep{van04}. But the 
region also contains considerable amounts of \ion{H}{2}
and possibly molecular gases (e.g., \citet{shu04}), in which 
one expect a strong upstream bow shock. 
We assume that \xs\ is located in the vicinity of \sgra. The
absorption-corrected luminosity of \xs\ is then $L_x \sim 1.0 \times
10^{34}$~erg~s$^{-1}$ in the 2--10 keV band,
well within the luminosity range of similar PWNe such as B1929+10
\citep{wan93, bec05}, N157B \citep{wan01}, 
and several more recently discovered nebulae (\citet{gae04} and references therein).
In particular, the Mouse nebula (powered by PSR~J1747--2958; \citet{gae04})
is very similar to \xs\ in their X-ray morphologies,
spectral steepening behaviors, and luminosities.

We can estimate the spin-down
power $\dot E$ of the putative pulsar in \xs\
from the empirical relationship $\Gamma_{PWN} \approx
2.4-0.66 \dot{E_{37}}^{-1/2}$ of \citet{got03}. This applies to all
rotation-powered known pulsars with observed bright PWNe, of which
\xs\ apparently qualifies. Based on our measured mean power law
index ($\Gamma_{PWN} \sim 1.9$) for \xs\ we estimate a pulsar
spin-down power of $\dot E \sim 3 \times 10^{37} {\rm
erg~s}^{-1}$. But with a considerable dispersion about the above empirical 
relationship \citep{got03}, this estimate is uncertain by at least 
a factor of 10.

In the ram-pressure confined PWN scenario, the pulsar
is located inside the leading edge.  The bulk of the pulsar spin-down energy is expected to be released in a
relativistic wind consisting of particles (assumed to 
be mainly electrons and positrons). A terminal 
shock should thermalize the bulk of pulsar wind particles 
to achieve a more-or-less Maxwellian distribution with the peak at $\gamma 
\sim \gamma_w$, the Lorentz factor of the pulsar wind (e.g., \citet{aro94}). 
The particle acceleration at the shock modifies this distribution to form
a high energy power law component, which is traced by the observed X-ray 
synchrotron emission. A particle that generates the synchrotron 
emission with a characteristic photon energy $\epsilon$ typically 
has a Lorentz factor 
\begin{equation}
\gamma_x \sim 3 \times 10^7 (\epsilon/H_{-4})^{1/2}, 
\label{eq:gamma}
\end{equation}
where $H_{-4}$ is the magnetic
field in units of $10^{-4}$~G.  

In a ram-pressure confined PWN, the shocked wind material is pushed
downstream into a tail in the opposite direction of the pulsar motion. 
With a large ambient pressure gradient and no significant gas entraining,
the flow can reach a velocity ($v_f$) comparable to the sound speed of the 
ultra-relativistically hot plasma,  
$c_s = c/\sqrt{3}$ \citep{van88, wan93, buc05}. The time for the flow to pass through the 
observed length of \xs\ ($l_x \sim 8^{\prime\prime} = 0.3$~pc) is
$t_x \sim l_x/(v_f \sin i) \sim (1.7{\rm~yrs})/\sin i, $
where $i$ is the inclination angle of the flow relative to the line of sight. 
To account for the observed spectral steepening along \xs,
this flow time  should be roughly equal to the X-ray synchrotron lifetime 
of the particles, $t_s \sim (40 {\rm~yrs}) 
\epsilon^{-1/2} H_{-4}^{-3/2}$, assuming no further acceleration of the
 downstream particles. Therefore, the magnetic field is
\begin{equation}
H \sim (8 \times 10^{-4} {\rm~G}) (\sin i)^{2/3} \epsilon^{-1/3} (v_f/c_s)^{2/3}.
\label{eq:v_f}
\end{equation}
The total magnetic field energy can then be estimated as
\begin{equation}
W_H \approx {H^2 \over 8\pi} V_x \sim (5 \times 10^{45} {\rm ergs}) (\sin i)^{1/3} \epsilon^{-2/3}(v_f/c_s)^{4/3},
\label{eq:ep}
\end{equation}
where $V_x \sim (2\times 10^{53} {\rm~cm^{3}})/\sin i$ is the 
total X-ray-emitting volume of \xs, assumed to have an approximate prolate 
shape with the major and minor axis to be 8\as\ and 2\as\ (\S~3). 
This energy is comparable to $1/2  t_x \dot E$ within the
uncertainties of the parameter values (e.g., with 
$\epsilon \sim 2$~keV, $\dot{E}_{37} \approx 3$, 
$\sin i \sim 0.7$, and $v_f \sim c_s$), indicating that the input energy 
from the pulsar is reasonably equal-partitioned between the thermal 
and magnetic field forms in \xs. 

From the observed X-ray spectral properties of \xs, we can further
infer the underlying energy distribution of the synchrotron particles. 
We assume that freshly accelerated particles have a differential energy
distribution 
\begin{equation}
N(\gamma) \propto \gamma^{-p}. 
\label{eq:ngamma1}
\end{equation}
From the observed power law 
photon index $\Gamma \sim 1.4$ (Fig.~\ref{prof}b) at the head of \xs, we obtain $p = 2 
\Gamma-1 \sim 1.8$. Because of the synchrotron cooling, the
higher energy part of the distribution steepens in the downstream. 
Over the entire PWN, the balance between the acceleration and the
synchrotron cooling leads to an energy distribution of accumulated particles 
\begin{equation}
N(\gamma) \propto \gamma^{-(p+1)}.
\label{eq:ngamma2}
\end{equation}
The break  between Eqs.~\ref{eq:ngamma1} and \ref{eq:ngamma2} occurs
at  $\gamma_b \sim 1 \times 10^6 t_3^{-1} H_{-4}^{-2}$, where $t_3$ is the age of the
pulsar in units of $10^3$~yrs \citep{che00}. 
The expected steepening ($\delta\Gamma = 0.5$) in the 
corresponding synchrotron spectrum is consistent with the 
value  for the entire \xs\ 
[$\Gamma \sim 1.94(1.80, 2.11)$, corresponding to 
$p=1.9(1.6, 2.2)$; \S~3]. 
At lower energies where the synchrotron cooling does not dominate,
other processes become important, such as 
inverse Compton scattering (ICS) in the nebula.
The corresponding particle energy distribution can be probed through 
observations of synchrotron radiation at lower photon energies and  
$\gamma$-rays from the ICS (\S~5).

No radio counterpart is found for \xs\ 
(Fig.~\ref{ims}b). The point-like radio emission from the putative pulsar may be too faint
to be present in existing observations and may be beamed away from our 
direction. The absence of the  extended radio emission 
indicates the lack of particles with $\gamma \lesssim 10^4$ 
(see Eq.~\ref{eq:gamma}), presumably 
because the pulsar wind has a substantially greater $\gamma_w$. 
Fig.~\ref{ims}b does show a possibly related radio enhancement 
just south of IRS 13.
This enhancement does not seem to be associated with any stellar
complex and shows an unusually low radio recombination to continuum ratio
(Shukla, Yun,  \& Scoville 2004). The enhancement seems to be aligned well
with the X-ray-emitting \xs\ and thus may represent the terminal shock of the 
downstream flow. We do not expect a significant 
accumulation of the radio-emitting synchrotron particles in the PWN, however,
because of their efficient ICS cooling in the strong ambient photon field (\S~5). 

\section{G359.95-0.04 as a TeV Source}

The presence of a young energetic pulsar, only 8\farcs7 from \sgra, can
also explain the very high energy (VHE; $\gtrsim 0.1$ TeV)
$\gamma$-ray radiation observed from the GC (e.g., \citet{aha04, alb05}). The HESS array of
Cherenkov telescopes, in particular, gives an  emission 
centroid error radius of $\sim 1^\prime$, which does not allow for a
spatial separation between \xs\ and \sgra. 
Various mechanisms for generating the emission have been proposed, almost
all of which are related to \sgra, Sgr~A~East, or the hypothetical 
annihilation of
super-symmetric dark matter particles (e.g., \citet{ahaN05} and
references therein). In addition, \citet{qua05} 
have suggested that stellar wind shocks may efficiently accelerate
particles to relativistic energies, which might scatter
ambient photons to TeV energies. However, 
there is little observational evidence for the expected diffuse
(synchrotron) X-ray emission from such particles, except from \xs, in the 
vicinity of Sgr A$^*$. In addition, younger and more massive stellar 
clusters, such as Arches and
Quintuplet, are not detected as TeV sources in the HESS survey. 

G359 as a PWN gives a ready explanation for
the VHE $\gamma$-ray emission, as the well-known example of the Crab
Nebula shows. Interestingly, the only other HESS source within the inner
$3^{\circ} \times 3^{\circ}$ region of the GC is a known PWN in
SNR~G0.9+0.1 \citep{aha05}. An one-zone synchrotron
plus ICS model well describes the data from radio to VHE $\gamma$-ray
with reasonable physical parameters of the PWN. 
We note that  \citet{ato05} have proposed that the VHE
$\gamma$-ray radiation from the vicinity of Sgr A$^*$ 
may be produced by a hypothetic wind nebula, which could be produced
by the SMBH, in a way similar to a PWN.  \xs\ is about four times
more luminous than \sgra\ in its quiescent state \citep{bag03} 
and is also much more extended, consistent with the lack of the
variability among the HESS observations.
Therefore, \xs\ is the most logic suspect for 
the VHE $\gamma$-ray radiation.

In a PWN, VHE $\gamma$-ray radiation 
arises from the ICS of lower-energy ambient seed
photons (e.g., the cosmic microwave background and stellar light) by
energetic particles. Under the Thomson approximation, 
a seed photon with energy $\epsilon_o$ after an ICS reaches a characteristic 
 photon energy $\epsilon \sim \gamma^2 \epsilon_o$. This approximation is
good for $b= 4\gamma \epsilon_o/(m_e c^2) \lesssim 1$ \citep{mod05}, i.e., 
when the photon energy
in the rest frame of the particle is considerably smaller than $m_e c^2$.
Otherwise, the Compton recoil (discrete energy gain) by the particle must
be accounted for, resulting in the so-called Klein-Nishina (KN) effect, or the
reduction of the ICS efficiency. As a result, the ICS of 
ultraviolet photons, which are abundantly produced by massive GC stars, to the
TeV regime is strongly suppressed. Instead, the dominant ICS 
seed photon source for the TeV radiation in the vicinity of Sgr A$^*$ is 
the far infrared (FIR) emission (dust-reprocessed stellar light) with an 
energy density of $U = 10^{-8} U_{-8} {\rm~ergs~cm^{-3}}$
and a characteristic photon energy of $\epsilon_o \approx 0.04 $ eV 
\citep{dav92, qua05}. Therefore, we concentrate on the FIR-to-TeV ICS, 
assuming that \xs\ is located in the 
vicinity of Sgr A$^*$. 

To check the relative importance of the ICS and the 
synchrotron emission in particle cooling, we also need to 
consider the photon to magnetic energy density ratio $q \approx 25 U_{-8} 
H^{-2}_{-4}$. Following \citet{mod05},
we express the ratio of the ICS power ($P_i$) to the synchrotron 
power ($P_s$) as
\begin{equation}
P_i/P_s = qF_{KN}(b),
\label{eq:pratio}
\end{equation}
where the function $F_{KN}(b) \approx 1/(1+b)^{3/2}$
describes the KN suppression relative to 
the Thomson approximation. 
For the FIR-to-TeV ICS, the required Lorentz factor 
$\gamma_g \sim 5 \times 10^6$. The corresponding $F_{KN} \sim 0.2$ indicates 
only a moderate
KN suppression, which is further compensated by $q$.
The condition $qF_{KN}=1$ defines a characteristic particle Lorentz factor 
$\gamma_s \approx 3 \times 10^7 U_{-8}^{2/3}
H^{-4/3}_{-4} $ \citep{mod05}. For particles 
with $\gamma \ll \gamma_s$,  
($F_{KN} \sim 1$; the KN effect is not important), the ICS cooling is a 
factor of $\sim q$ more efficient than the synchrotron emission. 
As a result, the synchrotron cooling  diminishes from infrared to 
radio. For $\gamma \gg \gamma_s$, 
the ICS is strongly suppressed by the 
KN effect and the synchrotron radiation dominates.
For X-ray-emitting synchrotron particles, the ICS cooling may not
be totally negligible, largely depending on the actual value of $q$. 

We can estimate the TeV flux from the ICS, based on the observed X-ray 
synchrotron emission of \xs\ (\S~3). The  differential photon 
flux of the synchrotron emission 
as a function of $\gamma$ can be approximately expressed as
\begin{equation}
f_s(\gamma_g) = f_s(\gamma_x) (\gamma_x/\gamma_g)^{p},
\label{eq:fsys}
\end{equation}
where the index $p$ is chosen because both 
$\gamma_g$ and $\gamma_x$ are likely greater than $\gamma_b$ (\S 4). 
We further express the ratio of the ICS intensity $f_i$ to the 
synchrotron flux $f_s$ as

\begin{equation}
{f_i(\gamma_g) \over f_s(\gamma_g)} = {({d \gamma \over d \epsilon})_i P_i \over 
({d \gamma \over d \epsilon})_s P_s}
\label{eq:fics}
\end{equation}
where $({d \gamma \over d \epsilon})_i^{-1} \sim 8 \times 10^{-5} \gamma_g$ and $({d \gamma \over d \epsilon})_s^{-1} \sim 7 \times 10^{-15} H_{-4} \gamma_g$. Here we have neglected any Doppler effect that may be caused by the bulk 
motion of the flow. From Eqs~\ref{eq:pratio}, \ref{eq:fsys}, and \ref{eq:fics}
as well as $f_s(\gamma_x) \sim f_x$ at 1 keV (\S~3), we obtain
the expected flux at $\sim 1$ TeV as 
\begin{equation}
f_i(\gamma_g) \sim  9 \times 10^{-12} {\rm~ph/(TeV~s~cm^2)} U_{-8} H_{-4}^{-2}, 
\label{eq:}
\end{equation}
for $p \sim 2$.

We can now compare the above estimate with the observed
spectrum of the TeV emission, which can be characterized by a power law
\citep{aha04, alb05}:
$f_g = (2.9\pm0.6) \times 10^{-12} \epsilon_{TeV}^{-2.2 \pm 0.2} 
{\rm~ph/(TeV~s~cm^2)}$. By having $f_i \sim f_g$, we 
obtain the  corresponding mean magnetic field strength in 
the ICS region as $H \sim 2 \times 10^{-4}$ G, which is comparable to the 
value from the synchrotron modeling of the X-ray data 
(\S 4). The power law index is marginally greater than 
that of the observed X-ray spectrum, which may be a result of an
expected spectral break at $\sim 4$ TeV due to the KN suppression of 
the ICS by particles with $\gamma \gtrsim \gamma_s$. Therefore,
the ICS of GC ambient FIR photons by particles in the PWN provides 
a consistent interpretation of the TeV radiation from the GC. 

\section{Connection to Other Galactic Center Objects}

Within the centroid uncertainty of the \xs\ pulsar, we find a
near-IR object [OEG99]192 ($17^h45^m39^s.78$, $-29^\circ 0^{\prime}
19\farcs9$).  This object has a K magnitude of 12.8, with evidence for
variation \citep{ott99} and is likely to be an
early-type star at the GC. The object could be a binary companion of
the pulsar. In this case, the ambient medium must be relatively dense, 
because a binary is physically not expected to
travel at a speed as high as would be expected for an isolated pulsar.
A proper motion of [OEG99]192 at a few $\times 10^2$~km~s$^{-1}$ would
also have been noticed in existing monitoring observations of the GC.
Alternatively, the object is not related to the pulsar. The
probability that [OEG99]192 represents only a chance position
coincidence is not small; there are other 10 SIMBAD objects in a 3\as\
radius, for example.

What could we learn about the origin of \xs\ from its well-defined 
cometary shape and the apparent motion away from the the GC cluster? The mere 
presence of such a coherent feature in this dynamic
region is quite extraordinary, where strong winds (e.g., from Sgr A$^*$ and 
the GC cluster), or even the turbulent motion of hot gas, may be 
expected to have high velocities up to $\sim 10^3 {\rm~km~s}^{-1}$.
For the ram-pressure confinement to work, the relative motion
between the \xs\ pulsar and the ambient medium should have a velocity 
at least of the same order, unless the pulsar
happens to run into a relative dense cloud, which may be indicated by
the somewhat higher absorption toward \xs\ than toward other sources in the 
region (\S~3). Furthermore, to produce the observed orientation 
this velocity should be in the opposite direction of the \xs\ X-ray 
tail, i.e., in the similar  direction of
the expected outward winds. One possible way to generate such a high velocity is
that the \xs\ pulsar was ejected dynamically from the close vicinity of the
SMBH or an intermediate-mass black hole (IMBH; e.g., \citet{yu03}). 

One plausible site for the pulsar ejection from an IMBH is the IRS 13 complex.
\citet{mai04} have suggested that the complex is the remnant core 
of a massive star cluster and may contain an IMBH, which helps to bound the 
system from a complete disruption by the tidal field of the SMBH (see
also \citet{sch02}). As shown in \S~3, the comparable X-ray intensities 
and spectral signatures of IRS 13 and Sgr A$^*$ indicate 
that similar emission processes are involved. Thus, the putative IMBH may play
a role in the X-ray emission from IRS 13. 
With the present separation of $\sim 10$\as\ between IRS 13 and \xs, 
the required travel time of the pulsar is $\sim 4 \times
10^2$~yrs~$v_{p,3}^{-1}$, where $v_{p,3}$ is the proper motion of the 
pulsar (in units of $10^{3} {\rm~km~s}^{-1}$) relative to IRS 13. 
During this period of time,
the IRS 13 complex should be moving at $\sim 3 \times 10^2 {\rm~km~s}^{-1}$ 
under the gravitational pull of the SMBH with a mass of $\sim 3 \times 10^6
M_\odot$.  The complex may have
traveled a considerable fraction of the distance that the \xs\ has done after the 
ejection, which may 
explain a slight misalignment between IRS 13 and \xs. Alternatively,
this misalignment or the slight bending of the \xs\ tail to the west 
(Fig.~\ref{ims}) may be caused by the momentum impact of the winds from
the Sgr A$^*$ and the GC cluster. The \xs\ pulsar, if indeed ejected
dynamically from IRS 13, could then be
formed much earlier. This opens up a possible connection between the 
pulsar/IRS 13 and  SNR Sgr~A~East, which has an age of $\sim 10^4$ yrs 
\citep{par05}. The current separation between  the
explosion site (IRS 13)  and the SNR could be caused partly by the 
ambient ram-pressure inserted on the latter. 

It is also entirely possible
that \xs\ is not directly related to IRS~13 itself and/or the Sgr~A~East SNR.
But \xs\  may still likely originate in the GC cluster, to which
IRS~13 belongs, and may be responsible
for much of the diffuse and apparently thermal
X-ray-emitting structure in the vicinity of Sgr A$^*$ (\S~3). The small 
ionization parameter ($\sim 3 \times 10^{10}$~s~cm$^{-3}$; 
Table~\ref{tab:spec}) indicates that the gas was heated only recently.
The structure may represent heated ejecta of the SN
that produced the \xs\ pulsar. 
The evolution of this SNR could have been affected substantially 
by the winds from both young and old stars and from Sgr A$^*$ 
as well as by the large differential rotation and the strong magnetic
field that thread the ambient medium. 
It is, however, beyond the scope of the present work to carefully examine these
processes in the remnant evolution, let alone its interaction with the PWN. 
If \xs\ indeed results from a separate explosion, Sgr~A~East is then not the 
only SNR that has recently influenced the GC environment, hence the accretion
of the SMBH.

While the PWN interpretation of the \xs\ can be firmly established
only when a periodic signal from the pulsar is detected, observations
with fast timing capability are the natural next step.  Based on the
above results, we have estimated the exposure time needed to detect
putative pulsations from \xs\ using \chandra\ HRC
or  {\sl XMM-Newton} EPIC-pn observations acquired in the fastest 
imaging/timing mode. With the limited sensitivity or spatial resolution 
of these instruments, however, such an observation will need an exposure in excess 
of a mega-second. We thus strongly encourage follow-up timing
observations of \xs\ in radio, near-IR, or possibly $\gamma$-ray.

In conclusion, \xs\ as a ram-pressure confined PWN provides a natural
interpretation for the observed steady X-ray flux, comet-shape
morphology, featureless hard nonthermal spectrum, and spectral
steepening away from the putative pulsar. The presence of this PWN may
also be responsible for much of the enhanced diffuse X-ray emission
and the VHE $\gamma$-ray source in the vicinity of Sgr A$^*$.  An
understanding of this energetic nonthermal X-ray feature, whatever its
true nature may be, is important for studying the nuclear environment
of the Galaxy.

\acknowledgments
We thank R. Genzel for providing the SINFONI near-IR image, 
M. Muno, R. Sari, and P. Goldreich for valuable comments, 
and the anonymous referee for useful suggestions that led to 
various improvements in the paper. This work is supported in part by 
SAO/NASA under grant SAO GO4-5010X. QDW is grateful to the hospitality that
he received at the Institute for Advanced Study, where much of this work 
was completed. FJL is partially supported by the Natural Science Foundation of China.

{}


\begin{thebibliography}{}
\bibitem[Aharonian et al.  (2004)]{aha04}Aharonian, F., et al.  2004, A\&AL, 425, 13
\bibitem[Aharonian et al.  (2005)]{aha05}Aharonian, F., et al. 2005, A\&AL, 432, 25
\bibitem[Aharonian \& Neronov (2005)]{ahaN05}Aharonian, F., \& Neronov, A. 2005, ApJ, 619, 306
\bibitem[Albert et al. (2005)]{alb05} Albert, J., et al, 2005, ApJL, submitted (astro-ph/0512469)
\bibitem[Arons \& Tavani (1994)]{aro94}Arons, J., \& Tavani, M. 1994, ApJS, 90, 797
\bibitem[Atoyan \& Dermer (2005)]{ato05}Atoyan, A., \& Dermer, C. D. 2005, ApJL, 617, 123
\bibitem[Baganoff et al. (2003)]{bag03}Baganoff, F. K., et al. 2003, ApJ, 591, 891
\bibitem[Becker et al. (2005)]{bec05}Becker, W., et al. 2005, ApJ, submitted, (astro-ph/0506545)
\bibitem[Belanger et al. (2005)]{bel05}Belanger, G., et al. 2005, ApJ, in press (astro-ph/0508128)
\bibitem[Bucciantini, Amato, \& Del Zanna (2005)]{buc05}Bucciantini, N.,  Amato, E., \& Del Zanna, L. 2005, A\&A, 434, 189
\bibitem[Chevalier (2000)]{che00}Chevalier, R. A. 2000, ApJL, 539, 45
\bibitem[Coker, Pittard, \& Kastner (2002)]{cok02}Coker, R. F., Pittard, J. M., \& Kastner, J. H. 2002, A\&A, 383, 568
\bibitem[Cordes \& Lazio (1997)]{cor97} Cordes, J. M., \& Lazio, T. J. 1997, ApJ, 475, 557
\bibitem[Davidson et al. (1992)]{dav92} Davidson, J. A., et al. 1992, ApJ, 387, 189
\bibitem[Eisenhauer et al. (2005)]{eis05}Eisenhauer, F., et al. 2005, ApJ, 628, 246
\bibitem[Gaensler et al. (2004)]{gae04}Gaensler, B. M., et al. 2004, ApJ, 616, 383
\bibitem[Genzel et al. (2003)]{gen03}Genzel, R., et al. 2003, ApJ, 594, 812
\bibitem[Gotthelf (2003)]{got03}Gotthelf, E. V. 2003, ApJ, 291, 361
\bibitem[Ho et al. (1985)]{ho85}Ho, P. T. P., Jackson, J. M., Barrett, A. H., \& Armstrong, J. T. 1985, ApJ, 288, 575
\bibitem[Krabbe et al. (1995)]{kra95}Krabbe, A., et al. 1995, ApJL, 447, 95
\bibitem[Lu, Wang, \& Lang (2003)]{lu03}Lu, F. J., Wang, Q. D., \& Lang, C. C. 2003, AJ, 126, 319
\bibitem[Maillard et al. (2004)]{mai04}Maillard, J. P., Paumard, T., Stolovy, S. R., \& Rigaut, F. 2004, A\&A, 423, 155 
\bibitem[Melia \& Falcke (2001)]{meli01}Melia, F., \& Falcke, H.  2001, ARA\&A, 39, 309
\bibitem[Moderski et al. (2005)]{mod05}Moderski, R., Sikora, M., Coppi, P. S., \& Aharonian, F. 2005, MNRAS, 363. 954
\bibitem[Muno et al. (2003)]{mun03}Muno, M. P., et al. 2003, ApJ, 589, 225
\bibitem[Muno et al. (2005)]{mun05}Muno, M. P., et al. 2005, ApJ, 633, 228
\bibitem[Ott, Eckart, \& Genzel (1999)]{ott99}Ott, T., Eckart, A., \& Genzel, R. 1999, ApJ, 523, 248
\bibitem[Park et al. (2005)]{par05}Park, S., et al. 2005, ApJ, 631, 964
\bibitem[Quataert \& Loeb (2005)]{qua05}Quataert, E., \& Loeb, A. 2005, ApJ, submitted (astro-ph/0509265)
\bibitem[Reid et al. (1999)]{rei99} Reid, M. J., Readhead, A. C. S., Vermeulen, R. C., \& Treuhaft, R. N. 1999, ApJ, 524, 816
\bibitem[Roberts \& Goss (1993)]{rob93}Roberts, D. A., \& Goss, W. M. 1993, ApJS, 86, 133
\bibitem[Sakano et al. (2003)]{sak03}Sakano, M., Warwick, R. S., Decourchelle, A., \& Predehl, P. 2003, MNRAS, 340, 747
\bibitem[Sch\"odel et al. (2005)]{sch02}Sch\"odel, R., et al. 2002, Nature, 419, 694
\bibitem[Shukla, Yun, \& Scoville (2004)]{shu04}Shukla, H., Yun, M. S., \& Scoville, N. Z. 2004, ApJ, 616, 231
\bibitem[Tan \&  Draine  (2004)]{tan04}Tan, J. C., \&  Draine, B. T.  2004, ApJ, 606, 296
\bibitem[van Buren \& McCray (1988)]{van88}van Buren, D., \& McCray, R.  1988, ApJL, 329, 93
\bibitem[van der Swaluw, Downes, \& Keegan  (2004)]{van04}van der Swaluw, E., Downes, T. P., \& Keegan, R. 2004, A\&A, 420, 937
\bibitem[Wang et al. (1993)]{wan93}Wang, Q. D., Li, Z.-Y., \& Begelman, M. C. 1993, Nature, 364, 127
\bibitem[Wang \etal\ (2001)]{wan01}Wang, Q. D., Gotthelf, E. V., Chu, Y.-H., \& Dickel, J. R. 2001, ApJ, 559, 275
\bibitem[Wang, Gotthelf, \& Lang (2002)]{wan02a}Wang, Q. D., Gotthelf, E. V., \& Lang, C. C. 2002, Nature, 415, 148
\bibitem[Wang, Lu, \& Lang (2002)]{wan02}Wang, Q. D., Lu, F. J., \& Lang, C. C. 2002, ApJ, 581, 1148
\bibitem[Xu et al. (2005)]{xu05}Xu, Y. D., et al. 2005, ApJ, in press (Astro-ph 0511590)
\bibitem[Yu \& Tremaine (2003)]{yu03}Yu Q., \& Tremaine S. 2003, ApJ, 599, 1129
\end{thebibliography}
\end{document}